\documentclass[preprint,showpacs,superscriptaddress,floatfix]{revtex4}
\usepackage{graphicx}
\usepackage{dcolumn}
\usepackage{bm}
\usepackage{color}

\newcommand{\figwid}{10cm}
\newcommand{\tfigwid}{6cm}
\newcommand{\cm}{cm$^{-1}$}

\newcommand{\bb}{K$_{0.3}$MoO$_{3}$}
\newcommand{\rb}{Rb$_{0.3}$MoO$_{3}$}
\newcommand{\ab}{A$_{0.3}$MoO$_{3}$}

\newcommand{\tcdw}{T$_{CDW}$}

\begin{document}
\title{A Raman study of the Charge-Density-Wave State in \ab\ (A = K,Rb)}
\author{D.~M. Sagar\footnote{\emph{Present address:}
Department of Chemistry, University of McGill, 801 Sherbrooke St.West, Montreal, QC, H3A2K6, CANADA.}}
\author{D. Fausti}
\affiliation{Zernike Institute for Advanced Materials, University of Groningen, 9747
AG Groningen, The Netherlands.}
\author{S. Yue}
\author{C.~A. Kuntscher}
\affiliation{Experimentalphysik 2, Universit\"{a}t Augsburg, 86135
Augsburg, Germany.}
\author{S. van Smaalen}
\affiliation{Laboratory of Crystallography, University of Bayreuth,
95440 Bayreuth, Germany.}
\author{P.~H.~M. van~Loosdrecht}\email{p.h.m.van.loosdrecht@rug.nl}
\affiliation{Zernike Institute for Advanced Materials, University of Groningen, 9747
AG Groningen, The Netherlands.}

\date{\today}
\begin{abstract}
We report a comparative Raman spectroscopic study
of the quasi-one-dimensional charge-density-wave systems \ab\ (A = K,
Rb). The temperature and polarization dependent experiments reveal
charge-coupled vibrational Raman features. The strongly
temperature-dependent collective amplitudon mode in both materials
differ by about 3~\cm, thus revealing the role of alkali atom. We
discus the observed vibrational features in terms of
charge-density-wave ground state accompanied by change in the
crystal symmetry. A frequency-kink in some modes seen in \bb\ between
T = 80 K and 100 K supports the first-order lock-in transition,
unlike \rb. The unusually sharp Raman lines(limited by the
instrumental response) at very low temperatures and their
temperature evolution suggests that the decay of the low energy
phonons is strongly influenced by the presence of the temperature
dependent charge density wave gap.
\end{abstract}
\pacs{
71.45.Lr Charge-density-wave systems, 
71.30.+h Metal-insulator transition and other electronic
transitions, 78.30.-j Infrared and Raman spectra
}
\maketitle
\section{introduction}
Quasi-one dimensional (1D) metals are interesting materials because
of their tendency to undergo a phase-transition inevitably due to
the inherent instability called the Peierls instability
\cite{Peierls55}. This instability, originating from the
electron-phonon interaction leads to a metal-insulator transition at
reduced temperature, where the insulating state may be characterized
by a lattice deformation in conjunction with a charge modulation and
the opening of a single particle excitation gap in the electronic
spectrum.\cite{Gruner94} One of the striking features of these
materials are their unusual non-linear electrical conduction
properties \cite{Gruner88,Tessema87,Zettl82}, which arises due to
the collective motion of the charge-density wave (CDW). Along with
the conventional single particle excitations, these materials also
have collective excitations, \textit{viz.}, amplitudons and phasons,
which dominate the low energy physical properties \cite{Gruner88}.
The Raman active collective amplitudon mode, \textit{i.e.} the
transverse oscillation of the amplitude of the
coupled charge-lattice wave, has been observed by various
researchers \cite{Travaglini83,Hirata01}.

Blue bronze, \ab\ (A = alkali metal) is one of the best known
quasi-1D materials which undergoes a CDW
transition\cite{Travaglini83,Gruner88,Perlstein68} via the Peierls
channel. The first evidence of the CDW transition in \bb\ by Raman
measurements came from the experiments by Travaglini and
Wachter\cite{Travaglini83}. 
In this paper a mode around 50 cm$^{-1}$\ was
observed at temperatures below \tcdw~$=180$~K that
showed a strong temperature dependence of, in
particular, its frequency when the temperature
approached TCDW from below.
As it was
also found that this is a fully symmetric ($A_{g}$-mode) breathing
mode it was assigned to the amplitude oscillation of the charge
density wave. Further evidence came from the observation that the
line width tends to diverge as the $T_{CDW}$ was approached from
below. This anomalous behaviour of the line width could be due to the
fluctuation effects that become pronounced in the vicinity of the
phase transition. Mean-field theory of a generalized second-order
phase transition predicts that, due to a temperature dependent order
parameter, the frequency (and the amplitude) of the soft mode tends
to zero as the temperature is approaches \tcdw.\cite{Gruner94}
However, in the the study by Travaglini \textit{et al}.
\cite{Travaglini83} the softening of the amplitudon mode with increasing
temperature was found to be incomplete. The observed softening of
only $\sim$13~\% was ascribed to 3D-correlations. The
3D-correlations are nothing but the long range Coulomb interaction
between density-modulated electrons, which might effectively screen
the electron-phonon interaction \cite{Becca96,Sykora05}, thereby
reducing the expected complete mode-softening. The effects of strong
local electron-electron interaction leads to both dynamical and
static screening processes dressing the electron-phonon coupling.
These screening effects are thought to be a generic feature of
strongly correlated electron systems.

Recently a Raman study of the anomalous profile of the amplitudon mode in \bb\ 
was reported \cite{Nishio00} in which for temperatures below 100 K the
amplitudon mode showed an anomalous "fin"-like profile toward the Stoke's side of
it's central frequency. This was interpreted as splitting of the amplitudon mode
into two or more closely spaced harmonic oscillator type modes. The
splitting was claimed to be due to a strong perturbation of the
charge density wave by impurity potentials. At very low temperatures
where there are practically no thermally excited electrons the
impurity potentials remain unscreened, leading to a spatial
variation of the CDW amplitude and a splitting of the amplitudon mode.

The present paper revisits the vibrational and electronic
excitations in the blue bronzes through a comparative study of \rb\ 
and \bb\ in order to examine the behaviour of the amplitudon mode in both the
compounds together with a detailed temperature and polarization
dependent study of the low energy phonons.

\section{Experimental details}
At room temperature the \bb\ and \rb\ bronzes crystallize in the
monoclinic C\textit{2/m} space group \cite{Ghedira85} with
lattice constants \textit{a}=16.23 {\AA}, \textit{b}=7.5502 {\AA},
\textit{c}=9.8 {\AA}, and $\beta$=94.89$^{0}$ for \bb\ and
\textit{a}=16.36 {\AA}, \textit{b}=7.555 {\AA}, \textit{c}=10.094
{\AA}, and $\beta$=93.87$^{0}$ for \rb. There are 20 formula units per unit cell.
The structure is build up by edge and corner sharing
clusters of ten strongly distorted MoO$_{6}$ octahedra
stacked along the [010] direction. These clusters
form layers in the (102) planes, which are separated by
the alkali ions.

At the charge density wave transition temperature the structure
transforms in an incommensurately modulated structure with a temperature
dependent incommensurate wave vector $\mathbf{q}_{ic}$ close to (but not equal to)
$0.25\mathbf{b}^\ast +\frac{1}{2}\mathbf{c}^\ast$.\cite{Schutte93,Fleming85}
At low temperature ($T<100$~K) the incommensurate wave vector becomes temperature independent,
even though this does not seem to correspond to a lock in transition, \textit{i.e.} the modulation
remains incommensurate.\cite{Fleming85} 

\bb\ and \rb\ single crystals with typical size
$5\times3\times0.5$~mm$^3$ were prepared by the temperature gradient
flux method\cite{Ram84}.
After polishing, samples were mounted in an
optical flow cryostat (temperature stabilization better than 0.1 K)
and backscattering Raman spectra were recorded (frequency resolution
$\sim$2~\cm) using a triple grating monochromator in subtractive
mode equipped with a liquid nitrogen cooled CCD detector. A
solid-state diode-pumped, frequency-doubled $Nd:YVO_{4}$\ laser
system is used as excitation source (wavelength 532~nm, spot size
10~$\mu$m). The power density was kept below 100~W/cm$^2$\ in order to
minimize heating effects. The polarization was controlled on both
the incoming and outgoing beam, giving access to all the
polarizations allowed for by the back-scattering configuration. In
general both parallel ( ($xx$), where $x$ is in the [20-1] direction,
and ($bb$) in Porto notation) and
perpendicular polarizations have been measured. The
perpendicular spectra did not yield appreciable scattering intensity
and will therefore not be discussed here.

\section{Raman spectra of \protect{K$_{0.3}$M\MakeLowercase{o}O$_{3}$} and {R\MakeLowercase{b}$_{0.3}$M\MakeLowercase{o}O$_{3}$} above and below \tcdw}

\begin{figure}[htb]
\includegraphics[width=\figwid]{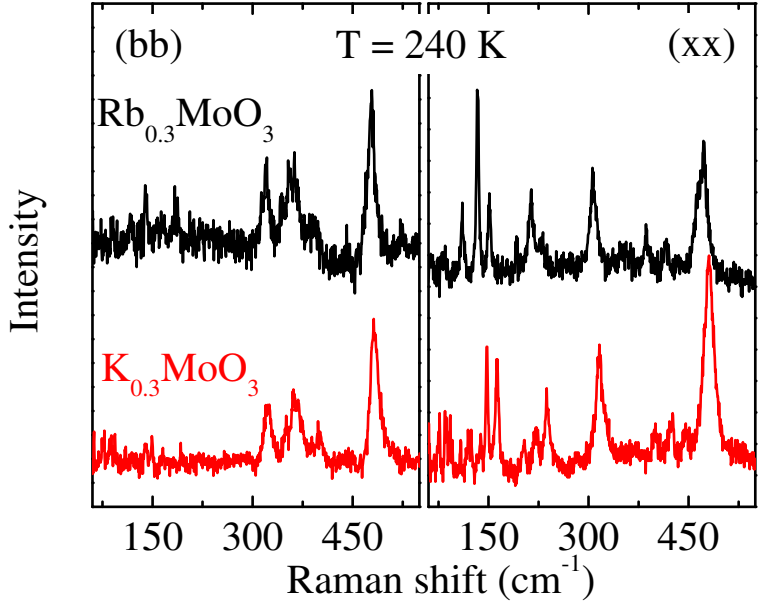}
 \caption{\label{ram1} 
 Polarized $(bb)$ and $(xx)$ Raman spectra of \bb\ and \rb\ above \tcdw.
 Spectra have been given an offset for clarity.}
\end{figure}

Fig. \ref{ram1} shows the Raman spectra of \bb\ and \rb\ above \tcdw\ 
for ($bb$) and ($xx$) polarization. Above \tcdw\ the Raman
spectra of \bb\ and \rb\ show very similar Raman vibrational features.
Only a few modes are observed in the ($bb$) spectra (Fig.~\ref{ram1}, left panel) 
since here the polarizations are along the metallic
direction. There is a strong mode near 475~\cm, a triplet with
frequencies near 320, 365, and 400~\cm, and some weaker modes below
200~\cm. The symmetric ($A_g$) modes above 300~\cm\ originate from
bending vibrations of the MoO$_6$ units, as has already been noted
in previous studies on blue bronzes\cite{Chua01,Pab05}, and is
consistent with the interpretation of the Raman spectra of pure
$MoO_{3}$\cite{Negishi04}. There are quite a few different modes
originating from the MoO$_6$ units, due to the large distortion of
the MoO$_6$ "octahedra"; the metal-oxygen distances in blue bronzes
typically range from 1.7 to 2.3~\AA \cite{Schutte93}. The
Mo-O stretching modes appear at higher frequencies (between 900 to
1000~\cm) \cite{Negishi04} as reported by Shigeru \textit{et al.,}
\cite{Nishio01} and Massa \cite{Massa90}. In general, the
frequencies for the modes observed in \rb\ are somewhat lower 
than the corresponding modes in \bb. This is consistent with the fact
that \rb\ has slightly larger lattice constants in the $a$\ and $c$\ 
directions\cite{Schutte93,Wang05} resulting from the larger ionic
radius of the rubidium ion. A small variance in the various Mo-O
distances between \bb\ and \rb\ has also been observed in the study of
Schutte \textit{et al.,} \cite{Schutte93}.

The right panel of Fig.\ref{ram1} shows the Raman spectra of \bb\ and
\rb\ ($T>$\tcdw) for the polarizations of the incident and scattered
light in the $x$ direction, \textit{i.e.} perpendicular to metallic
direction. As expected, the scattering efficiency in this direction
is substantially higher than in the metallic direction, and quite a
few more modes are observed. In both \rb\ and \bb\ there are about 20
modes that are resolved in the present experiment. Again, the
overall spectra are qualitatively very similar and many of the \rb\ 
vibrational modes are shifted toward slightly lower frequency in
comparison to the \bb\ modes. For instance the triplet near 150~\cm\ 
in the \bb\ bronze are found some 10-15~\cm\ lower in energy in the
\rb\ spectra, and the 480~\cm\ \bb\ mode is observed at 473~\cm\ in
\rb\ .

Finally, note that group theory predicts 71 A$_{g}$ modes for the
high temperature phase of the bronzes. The experimental spectra,
however, reveal only 35 modes. The "missing" modes most likely
result from accidental degeneracies and from weak intensity and/or
screening, in particular in the metallic direction. For the
predicted 55 B$_g$ modes (($xb$) geometry) the intensity is to weak
to be detected in the present experiments.

\begin{figure}[htb]
\includegraphics[width=\figwid]{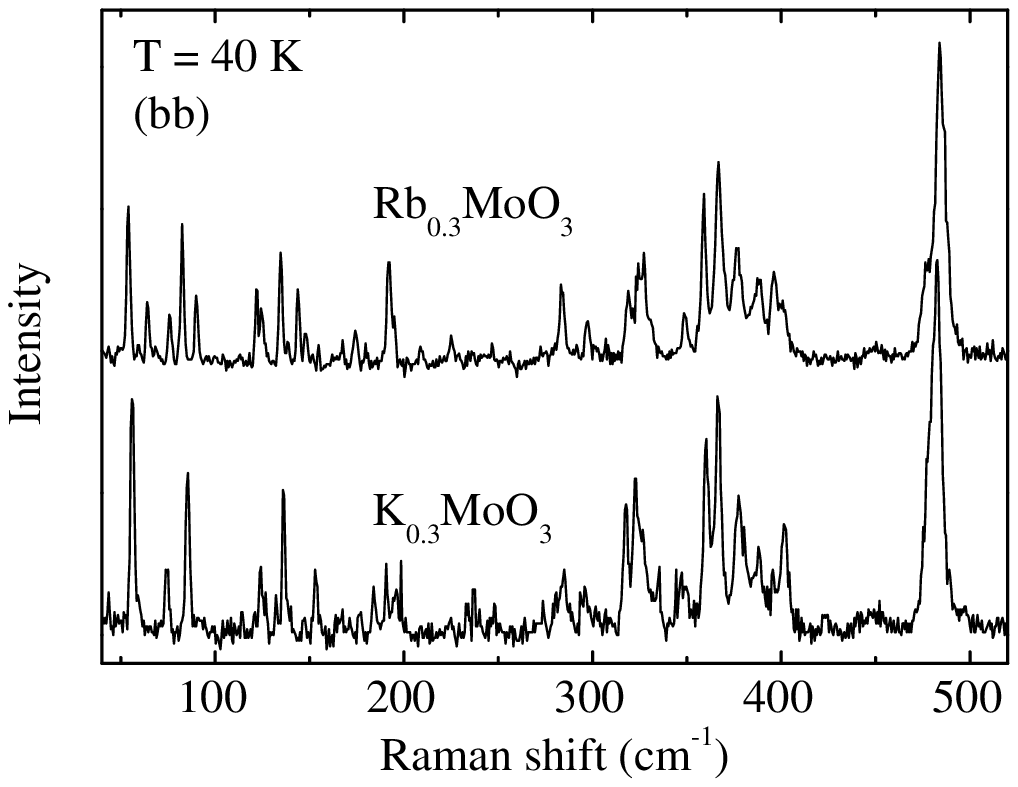}
 \caption{\label{ram4} ($bb$) Polarized Raman spectra of \bb\ and \rb\ below \tcdw.
 Spectra have been given an offset for clarity.}
\end{figure}

\begin{figure}[htb]
\includegraphics[width=\figwid]{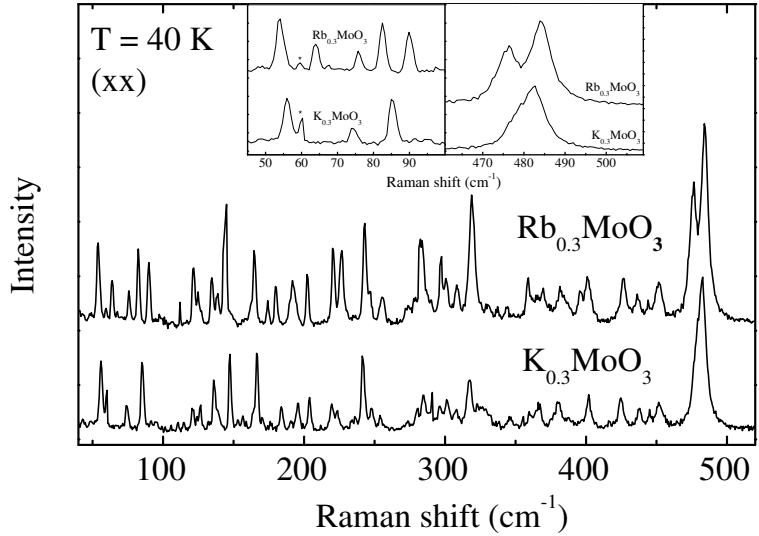}
 \caption{\label{ram3} ($xx$) Polarized Raman spectra of \bb\ and \rb\ below \tcdw. 
 Spectra have been given an offset for clarity.
 The insets show details of the splitting of some of the modes. The peak 
 at 60~\cm\ marked by an asterix is an experimental artefact originating from air.}
\end{figure}

The low temperature phase shows a much richer spectrum than the high
temperature one (see Figs. \ref{ram4} and \ref{ram3}), showing
many quite narrow Raman modes. The ($bb$) polarized low temperature
spectra (Fig. \ref{ram4}) exhibit 32 distinct modes for the \bb\ 
compound, and 42 in the \rb\ compound. For the ($xx$) polarization
(Fig. \ref{ram3}) the number of observed modes is 38 and 44 for \bb\ 
and \rb, respectively. Again, the spectra for the two compounds are
quite similar. The main difference is the number of observed modes
which seems to originate from an additional splitting of modes in
the \rb\ case. For instance the doublet observed near 75~\cm\ in \bb\ 
is seen as a triplet (see also left inset Fig.\ref{ram3}). We note
that, in a previous Raman study \cite{Massa90} a band near 80
\cm\ was observed in \rb\ which was ascribed to a zone-folded mode
arising from the folding of the original Brillouin-zone due to CDW
transition. The picture of zone-folding-active phonons is
corroborated by previous neutron scattering studies by Pouget
\textit{et al.,} \cite{Pouget91} in which a flat dispersion-less
phonon branch was observed. In the present study three modes are
observed in the \bb\ compound in the 40-100~\cm\ region, and five in
the \rb\ compound. As is discussed in section \ref{tdep}, the lowest
modes at 56~\cm\ (\bb) and 53~\cm\ (\rb) are the amplitudon mode of
the CDW. Note that only a single mode is observed here, and no
evidence is found for the "fin" like structure as discussed by
Shigeru \textit{et al.} \cite{Nishio00}. Also the Mo-O bending modes
show an additional splitting in the the \rb\ compound, which is
exemplified by the right inset of Fig. \ref{ram3} for the 480~\cm\
mode. As a general rule, splitting of Raman modes occurs mainly due
to two mechanisms. The first one occurs when a static distortion
transforms equivalent atomic positions to inequivalent ones
\cite{Tenne99}. A second mechanism is the correlation field or
Davydov\cite{Davydov71,Tenne99,Fausti06} splitting that occurs due
to coupling of vibrations of molecular units at different equivalent
sites in the unit cell, and is sensitive to transitions involving
multiplication of the unit cell size (like for instance the Peierls
distortion). The incommensurate nature of the blue bronzes may also
activate vibrational modes in Raman spectra\cite{Cur87}. Modes with
wave vector {\bf k}$=n${\bf q} become active in the incommensurate
phase, where {\bf q} is the incommensurate modulation vector, and
their scattering strength depends strongly on the modulation
amplitude.

\section{Temperature dependent Raman spectra\label{tdep}}
\begin{figure}[htbp]
\includegraphics[width=\figwid,clip=true]{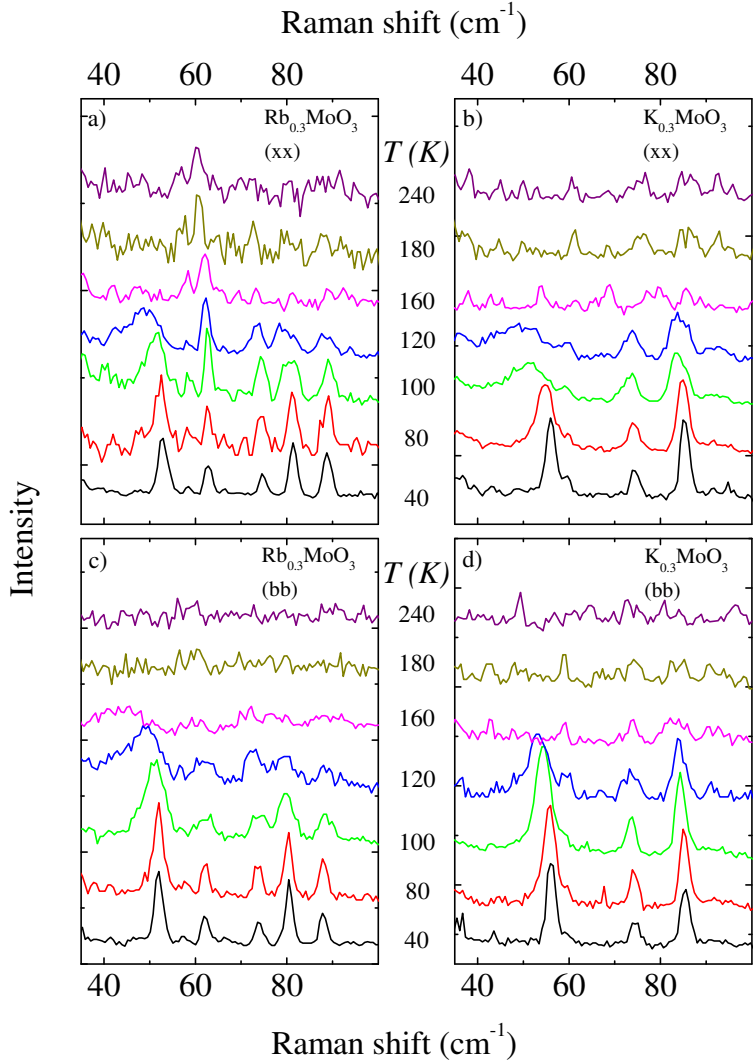}
 \caption{\label{ram45}
 Polarized Raman spectra of \bb\ and \rb\ at selected
 temperatures. Panel a) and b): ($xx$) spectra; Panel c) and
 d): ($bb$) spectra. The spectra have been given an offset for clarity.}
\end{figure}

The temperature dependence of the low frequency modes of
both compounds is depicted in Fig. \ref{ram45} for a few selected
temperatures. As mentioned before, in \bb\ there are only two clear phonons observed in
this region at 74~\cm, and 85~\cm, whereas the \rb\ compound shows four
phonon modes in this region centred at 64, 75, 83, and 90~\cm.
This, as well as the previously noted additional splitting in
some of the high frequency modes in \rb, strongly indicates that the
low temperature structures of the two compounds are not completely
identical. In general it is thought that the crystal structures of
\rb\ and \bb\ are, apart from the unit cell dimensions, the
same\cite{Schutte93,Ghedira85,Massa90}. The only relevant small
difference lies in the incommensurate modulation. Both structures
have their incommensurate modulation along the {\bf b$^\ast$}
direction. The amplitude of the modulation, however, is found to be
substantially larger for the \rb\ compound\cite{Schutte93}.
Therefore, we tentatively conclude that the observation of a larger
number of modes in \rb, as compared to \bb, is due to the larger
displacement amplitude of the incommensurate modulation in the low
temperature phase.

\begin{figure}[htb]
\includegraphics[width=\figwid,clip=true]{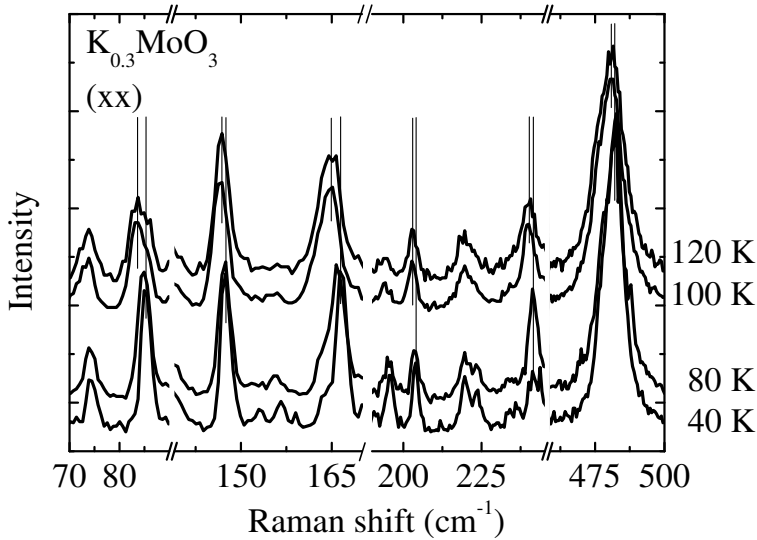}
 \caption{\label{ram7}
 Details of the ($xx$) polarized Raman spectra of \bb\ showing a discontinuous energy shift between 80K and 100K 
 for some of the modes. 
 The spectra have been given an offset for clarity.
 }
\end{figure}

Upon decreasing temperature the modes in \rb\ show a continuous behaviour,
and no further indications for a possible additional phase transition are
found. This in contrast to the situation in \bb\ . Upon lowering the
temperature, some of the modes show an abrupt change in their energy
in the between 80~K and 100~K. This is exemplified in Fig.~\ref{ram7}. Clearly some of
the modes show a discontinuous shift to higher energy of a few wave numbers indicating
a phase transition. It should be noted that not all modes exhibit the frequency-'kink',
as can for instance be seen for the 75~\cm\ mode in Fig.\ref{ram7}.
In general, a temperature-dependent frequency-jump is a signature of
a first-order transition. In the present case, it could be due to a lock-in
transition to a commensurate state. However, the low temperature phase of \bb\ 
has been studied in quite some detail and temperature dependent
neutron scattering experiments\cite{Fleming85}
failed to show a lock-in transition in \bb\ .
It was found that the incommensurate wave vector is strongly temperature
dependent down to $T = 100$~K. Below this temperature the wave vector
becomes temperature independent, but remains incommensurate.
Still, the present study strongly indicates that there is indeed a
first order phase transition in \bb\ below 100~K. 

The most salient feature in the low frequency spectra is
the appearance of the amplitudon in both compounds which shows a
softening and broadening upon approaching the phase transition (see
Fig.\ref{ram45}).
The low temperature frequencies of these modes are 56~\cm\ in the \bb\ 
compound, and 53~\cm\ in \rb\ . The interpretation of these modes
as amplitudon modes is consistent with
the previous neutron study by Pouget \textit{et al.,}
\cite{Pouget91}, and Raman scattering studies by Travaglini
\textit{et al.,} \cite{Travaglini83} and Massa \cite{Massa90}.
Recently this mode has also been observed in \bb\ as coherent
excitations in pump-probe transient reflectivity
experiments \cite{Demsar99,Sagar07}, in addition to coherent phonon
excitations\cite{Sagar07} at 74~\cm, and 85~\cm\ also observed
in the present data.
\begin{figure}[htb]
\includegraphics[width=\figwid,clip=true]{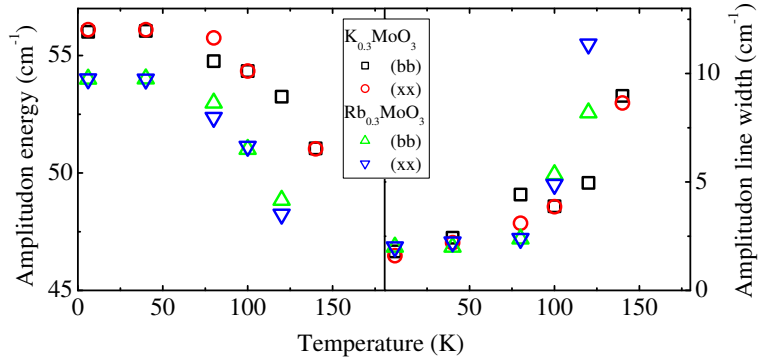}
 \caption{\label{linewidth}
 Temperature dependent energy (left panel) and line width (right panel) 
 of the amplitudon mode in \bb\ and \rb.
 }
\end{figure}
The amplitudon modes disappear from the spectra
already around 150~K. This is due to the strongly over damped nature of the amplitudon in
the vicinity of the phase transition to the metallic state originating from
strong fluctuations of the charge density wave. These fluctuations lead to
a diverging line width (see Fig. \ref{linewidth}, right panel),
which in turn makes it impossible to observe complete amplitude softening.
The observable amplitudon softening
upon approaching the phase transition amounts only
about 10 \%, as is shown in Fig. \ref{linewidth} (left panel).

\begin{figure}[htb]
\includegraphics[width=\tfigwid,clip=true]{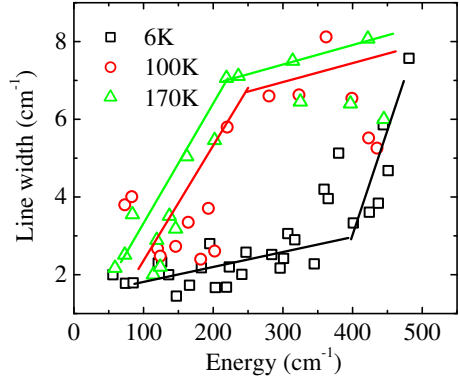}
 \caption{\label{ram11}
 Line width of the phonon modes in \bb\ as a function of their energy for
 $T=6$~K, 100~K, and 170~K.
 }
\end{figure}

Finally, turn back to the phonon modes observed in the CDW phase.
The phonon modes observed at low temperatures are rather narrow, and
often resolution limited. This is in particular true for the
phonons below 400~\cm.
Fig.\ref{ram11} displays the measured line width of the phonons
as a function of their frequencies in \bb\ for various temperatures.
As can be seen from the figure, at 6~K the line width of the
modes are very narrow up to $\sim$350~\cm. In contrast, the modes above
350~\cm\ show a progressive increase in their line width.
In other words, there is a sharp onset near 350~\cm in the
behaviour of the line widths. As temperature is increased this onset
shifts to lower frequencies. A similar behaviour is observed in \rb\ .

The observed anomalous line broadening hints to a coupling of the phonons to electronic
excitations with an energy above $\sim$350~\cm, which, given the temperature dependence,
scale with the CDW gap. Since the CDW energy gap in \bb\ is about 900~\cm\ it can not
be a coupling to the usual quasi-particle excitations.
It has been argued in the literature that a common feature of
incommensurate CDW materials is the presence of so called
midgap states\cite{Nakano86}. Therefore the anomalous broadening of the phonons
above $\sim$350~\cm\ is tentatively assigned to a coupling of the phonons to
midgap state excitations.

\section{Summary}
In summary, we have studied two representatives of the prototypical
quasi-one-dimensional charge density wave system, blue bronze, using
temperature and polarization dependent Raman scattering experiments.
The CDW transition lead to the activation of a large number of modes in the
low temperature phase of both the bronzes. The amplitudon mode, found at
nearly equal energy in \rb\ and \bb\ (there is a 2~\cm\ difference between them),
shows a rapid exponential broadening upon approaching the phase transition from
below making it impossible to observe a full softening of this mode.
The observation of a number of additional modes in the \rb\ compound
is assigned to the stronger incommensurate
modulation in this compound. This conclusion is, however, based on the limited
amount of low temperature structural studies, and it would be interesting
to perform additional studies to confirm the present interpretation.
This also holds for the observed changes in the \bb\ spectra between
80 and 100~K, which indicate a first order phase transition. This is in line 
with the recently discussion of the maximum in the threshold field for 
CDW conduction which was interpreted as due to a incommensurate-commensurate phase 
transition (lock-in transition).\cite{yue05} This transition is not 
related to the recently reported observation of 
a glass transition in \bb\ by Stare\`{s}ini\`{c} {\em et al.}\cite{Star04} at $T_g\sim10$~K, 
for which no further evidence was found in the present study.
Finally, evidence is found for a coupling of the high frequency phonons
to mid-gap excitations.

{\bf Acknowledgements }
This work is part of the research programme of the 'Stichting voor Fundamenteel 
Onderzoek der Materie (FOM)', which is financially supported by the 'Nederlandse 
Organisatie voor Wetenschappelijk Onderzoek (NWO)'.


\end{document}